\documentclass[longauth]{aa}
\usepackage[utf8]{inputenc}
\usepackage{linenoaa}
\usepackage[varg]{txfonts}
\usepackage{natbib}
\usepackage[normalem]{ulem}
\usepackage{array}
\usepackage[dvipsnames]{xcolor}
\bibpunct{(}{)}{;}{a}{}{,}
\newcommand{\Msun}{$\mathrm{M}_\odot$\xspace}

\begin{document} 

   \title{The nearby He-rich superluminous supernova SN~2021bnw during photospheric phases}

   \subtitle{}
\author{
A.~Fiore
\inst{1,2}\thanks{Corresponding author: achille.fiore@inaf.it} 
\and
A.~Kozyreva
\inst{3}
\and
L.~Yan
\inst{4}
\and
S.~Benetti
\inst{2}
\and
J.~P.~Anderson
\inst{5}
\and
P.~Baklanov
\inst{6,7,8}
\and
Y.~-Z.~Cai
\inst{2,9,10}
\and
E.~Cappellaro
\inst{2}
\and
T.-W.~Chen
\inst{11}
\and
N.~Elias-Rosa
\inst{2,12}
\and
A.~Gal-Yam
\inst{13}
\and
M.~J.~Graham
\inst{14}
\and
M.~Gromadzki
\inst{15}
\and
S.~L.~Groom
\inst{16}
\and
C.~P.~Guti{\'e}rrez
\inst{17,12}
\and
D.~Hiramatsu
\inst{18,19,20}
\and
D.~A.~Howell
\inst{21,22}
\and
C.~Inserra
\inst{23}
\and
M.~M. Kasliwal
\inst{14}
\and
R.~Könyves-T\'oth
\inst{24,25}
\and
P.~Lundqvist
\inst{26}
\and
C.~McCully
\inst{22}
\and
A.~Mironov
\inst{8}
\and
S.~Moran
\inst{27}
\and
T.~E.~M{ü}ller-Bravo
\inst{28,29}
\and
M.~Newsome
\inst{30}
\and
M.~Nicholl
\inst{31}
\and
P.~Ochner
\inst{2,32}
\and
E.~Padilla~Gonzalez
\inst{33}
\and
P.~J.~Pessi
\inst{26}
\and
G.~Pignata
\inst{34}
\and
F.~Ragosta
\inst{35,36}
\and
A.~Reguitti
\inst{2,37}
\and
T.~M.~Reynolds
\inst{38,39,40}
\and
R.~L.~Riddle
\inst{4}
\and
B.~Rusholme
\inst{16}
\and
I.~Salmaso
\inst{36,2}
\and
S.~Schulze
\inst{41}
\and
J.~Sollerman
\inst{26}
\and
L.~Tomasella
\inst{2}
\and
D.~Warshofsky
\inst{42}
\and
S. Yang
\inst{43}
\and
D.~R.~Young
\inst{31}
}
\institute{
INAF - Osservatorio Astronomico d’Abruzzo, Via Mentore Maggini Snc, 64100 Teramo, Italy
\and
INAF - Osservatorio Astronomico di Padova, Vicolo dell’Osservatorio 5, 35122 Padova, Italy
\and
Dipartimento di Fisica, “Sapienza” Università di Roma \& Sezione INFN Roma1, Piazzale Aldo Moro 5, 00185 Roma, Italy
\and
Caltech Optical Observatories, California Institute of Technology, Pasadena, CA 91125, USA
\and
European Southern Observatory, Alonso de C\'ordova 3107, Casilla 19, Santiago, Chile
\and
National Research Center, Kurchatov Institute, pl. Kurchatova 1, Moscow 123182, Russia
\and
Lebedev Physical Institute, Russian Academy of Sciences, 53 Leninsky Avenue, Moscow 119991, Russia
\and
M.V. Lomonosov Moscow State University, Sternberg Astronomical Institute, 119234 Moscow, Russia
\and
Yunnan Observatories, Chinese Academy of Sciences, Kunming 650216, P.R. China
\and
International Centre of Supernovae, Yunnan Key Laboratory, Kunming 650216, P.R. China
\and
Graduate Institute of Astronomy, National Central University, 300 Jhongda Road, 32001 Jhongli, Taiwan
\and
Institute of Space Sciences (ICE, CSIC), Campus UAB, Carrer de Can Magrans, s/n, E-08193 Barcelona, Spain
\and
Department of Particle Physics and Astrophysics, Weizmann Institute of Science, Rehovot, Israel, 76100
\and
Division of Physics, Mathematics and Astronomy, California Institute of Technology, Pasadena, CA 91125, USA
\and
Astronomical Observatory, University of Warsaw, Al. Ujazdowskie 4, 00-478 Warszawa, Poland
\and
IPAC, California Institute of Technology, 1200 E. California Blvd, Pasadena, CA 91125, USA
\and
Institut d'Estudis Espacials de Catalunya (IEEC), Edifici RDIT, Campus UPC, 08860 Castelldefels (Barcelona), Spain
\and
Department of Astronomy, University of Florida, 211 Bryant Space Science Center, Gainesville, FL 32611-2055 USA
\and
Center for Astrophysics | Harvard \& Smithsonian, 60 Garden Street, Cambridge, MA 02138-1516, USA
\and
The NSF AI Institute for Artificial Intelligence and Fundamental Interactions, USA
\and
Department of Physics, University of California, Santa Barbara, CA 93106-9530, USA
\and
Las Cumbres Observatory, 6740 Cortona Drive, Suite 102, Goleta, CA 93117-5575, USA
\and
Cardiff Hub for Astrophysics Research and Technology, School of Physics \& Astronomy, Cardiff University, Queens Buildings, The Parade, Cardiff, CF24 3AA, UK
\and
HUN-REN Research Centre for Astronomy and Earth Sciences, Konkoly Observatory, MTA Centre of Excellence, Konkoly Thege Miklós út 15-17., H-1121 Budapest, Hungary
\and
Department of Experimental Physics, Institute of Physics, University of Szeged, D\'om t\'er 9, Szeged, 6720 Hungary
\and
The Oskar Klein Centre, Department of Astronomy, AlbaNova, Stockholm University, SE-106 91 Stockholm, Sweden
\and
School of Physics and Astronomy, University of Leicester, University Road, Leicester LE1 7RH, UK
\and
School of Physics, Trinity College Dublin, The University of Dublin, Dublin 2, Ireland
\and
Instituto de Ciencias Exactas y Naturales (ICEN), Universidad Arturo Prat, Chile
\and
University of Texas at Austin, 1 University Station C1400, Austin, TX 78712-0259, USA
\and
Astrophysics Research Centre, School of Mathematics and Physics, Queen’s University Belfast, Belfast BT7 1NN, UK
\and
Department of Physics and Astronomy, University of Padova, Via F. Marzolo 8, I-35131 Padova, Italy
\and
Johns Hopkins University Bloomberg Center for Physics and Astronomy San Martin Dr, Baltimore, MD 21210
\and
Instituto de Alta Investigaci\'on, Universidad de Tarapac\'a, Casilla 7D, Arica, Chile
\and
Dipartimento di Fisica “Ettore Pancini”, Università di Napoli Federico II, Via Cinthia 9, 80126 Naples, Italy
\and
INAF - Osservatorio Astronomico di Capodimonte, Salita Moiariello 16, 80131 Napoli, Italy
\and
INAF - Osservatorio Astronomico di Brera, Via Bianchi 46, 23807 Merate (LC), Italy
\and
Tuorla Observatory, Department of Physics and Astronomy, University of Turku, FI-20014 Turku, Finland
\and
Cosmic Dawn Center (DAWN)
\and
Niels Bohr Institute, University of Copenhagen, Jagtvej 128, 2200 København N, Denmark
\and
Center for Interdisciplinary Exploration and Research in Astrophysics (CIERA), Northwestern University, 1800 Sherman Ave., Evanston, IL 60201, USA
\and
School of Physics and Astronomy, University of Minnesota, Minneapolis, MN55455, USA
\and
Henan Academy of Sciences, Zhengzhou 450046, Henan, China
}
   \date{Received September 15, 20XX; accepted March 16, 20XX}
  \abstract
   {}
   {We present and interpret the data of the nearby hydrogen-deficient but helium-rich superluminous supernova SN~2021bnw, which reached a magnitude of $-20.7$ at maximum luminosity in $g$ band.}
   {We discuss the light curves and spectra of SN~2021bnw based on its spectro-photometric follow-up, exploiting different observational facilities. We reproduced the near-IR spectrum of SN~2021bnw with \texttt{TARDIS} to inspect the chemical composition at late photospheric phases and identify helium features. We also used a \texttt{STELLA} model coupling hydrodynamics and radiation transport to constrain the physical parameters of the explosion assuming a $^{56}$Ni+CSM scenario.}
   {We suggest that SN~2021bnw was mainly powered by the interaction of the ejecta with a previously lost He-rich circumstellar material, coupled with a central power source.}
   {This work expands the data sample of He-rich superluminous supernovae (SLSNe Ib), and for a single-progenitor scenario, it can constrain the masses and physics of their progenitors.}
    \titlerunning{A nearby He-rich superluminous supernova in photospheric phases}
    \authorrunning{Fiore et al.}
   \maketitle
   \nolinenumbers
\section{Introduction}
\label{sec:intro}
At the end of their life, massive stars \citep[$\gtrsim8\,\mathrm{M_\odot}$, e.g.,][]{smartt2009} undergo gravitational collapse of their core, and their outer layers can be ejected in a catastrophic fashion. This event may be accompanied by a luminous supernova (SN), whose luminosity is usually comprised between $-14\,\mathrm{mag}$ and $-19$ at maximum in optical bands. A common explanation for the luminosity of stripped-envelope SNe \citep[][]{2018Taddia} considers the radioactive decay of  $\lesssim0.1\,\mathrm{M_\odot}$ of $^{56}$Ni on average (\citealt{rodriguezetal2023}; see also \citealt{anderson2019}). In the past two decades, observations of superluminous supernovae \citep[SLSNe;][]{galyam2019} questioned this scenario. It is challenging for the classical neutrino-driven core-collapse paradigm to synthesize more $^{56}$Ni (see, e.g., \citealt{uglianoetal2012,pejchaandthompson2015,sukhboldetal2016,terreranetal2017,suwaetal2019}, but see also \citealt{umedaandnomoto2008} for higher estimates of the $^{56}$Ni yield). The contribution of the spindown radiation from a newly born and rapidly spinning neutron star \citep[e.g.,][]{woosley2010,kasenandbildsten2010,nicholletal2017,suzukiandmaeda2019,dessart2019,suzukiandmaeda2021} and/or that of the interaction of the SN ejecta with a previously lost circumstellar material \citep[CSM; e.~g.][]{chevalierandfransson2003, chevalierandirwin2011} are viable explanations for interpreting SLSNe \citep[e.~g.][]{chatzopoulosetal2012,chatzopoulosetal2013,vreeswijketal2017,yanetal2015,yanetal2017,marguttietal2023}. Upon making reasonable assumptions to account for the lack of multicomponent features in the spectra of hydrogen-poor SLSNe \citep[see, e.g.,][]{smith2017}, CSM interaction also suitably explains the undulations often seen in their light curves (LCs), which are not naturally predicted by the magnetar scenario \citep[but see][]{moriyaetal2022,dongetal2023,zhuetal2024,zhangetal2025,farahetal2026}. For some events, a combination of these two mechanisms was suggested as powering mechanism \citep[e.g.,][]{fioreetal2021,hosseinzadehetal2022,chenetal2023b,westetal2023}. Still, the nature of SLSNe I remains elusive. Even though the increase in SLSNe-I data has enabled the community and the search for correlations among physical parameters derived from semi-analytic LC fits \citep{nicholletal2015,chenetal2023b,gomezetal2024}, firm conclusions have yet to be reached about their nature. It remains challenging to unambiguously identify the dominant power source for these events \citep[see, e.g.,][]{koenyvestoth2025}. In this regard, the interpretation of SLSN-I nebular spectroscopy offers a promising tool. A putative central magnetar is expected to affect the ionization and hydrodynamics of the ejecta at late epochs \citep{blanchardetal2026}, thereby leaving distinct signatures in the nebular spectra \citep{omandandjerkstrand2023,omand2026}. In addition, the detection of an early LC bump compatible with the breakout of a relativistic jet \citep[e.g.,][]{margalitetal2018, gottliebandmetzger2024} can also point to the contribution of a central compact object.
Another possibility would be that SLSNe are the result of a pair-instability SN (PISN) explosion, in which a very massive progenitor \citep[with a He-core mass  $64\,\mathrm{M}_\odot\lesssim M_{\rm He\,core}\lesssim 133\,\mathrm{M}_\odot$, ][]{hegerandwoosley2002} collapses due to the onset of $(e^{-},e^{+})$ pair production and allows for a massive synthesis of $^{56}\mathrm{Ni}$. This mechanism is capable of reproducing very bright and long-lived SNe in principle, with spectra suppressed on the blue side by the high opacities of the Fe-group elements \citep[e.g.,][]{mazzalietal2019}. For these reasons, some PISN candidates \citep[e.g.,][]{galyametal2009,nicholletal2013,tinyanontetal2023} have been reinterpreted under the assumption of the magnetar scenario \citep{nicholletal2013,chenetal2015,tinyanontetal2023}. Conversely, the PISN model has been invoked by \citet{schulzeetal2024} to explain the superluminous SN~2018ibb.

Observationally, SLSNe are classified as Type II and Type I SLSNe depending on whether their spectra show Balmer lines \citep{galyam2012}. In this work, we focus on hydrogen-poor events. Their LCs are very heterogeneous and continuously span between very slow or very fast evolving timescales \citep[see, e.g.,][]{deciaetal2018,inserra2019,chenetal2023a}. In addition, LCs of SLSNe I can show bumps before and/or after maximum whose amplitude is usually $<1$ mag \citep[e.g.,][]{nicholletal2015,nicholletal2016,inserraetal2017,lunnanetal2018,angusetal2019,fioreetal2021,gutierrezetal2022,hosseinzadehetal2022,linetal2023,chenetal2023a,westetal2023}. Moreover, spectra of SLSNe I around maximum are usually hot ($\sim15,000-20,000$ K) and often present prominent absorptions around $3000-5000$ $\text{\AA{}}$, which are usually interpreted as \ion{O}{ii} features (\citealp{quimbyetal2018}, but see \citealp[for an alternative explanation]{koenivestothetal2020})
. However, SLSNe I do not always exhibit these features \citep{koenyvestothetal2023}, but \citet{saitoetal2024} and \citet{nicholl2026} showed that their occurrence depends on their photospheric temperature. The spectra of SLSNe~I look remarkably similar to those of SNe Ic at maximum luminosity at $15-20$ rest-frame days after maximum  \citep[see, e.g.,][]{galyam2019,inserra2019}. This motivated the quest for a possible connection between SLSNe I and hypernovae \citep[see e.~g.][]{mazzalietal2016,modjazandbianco2016,nicholletal2017,fioreetal2025}. Finally, SLSNe~I hosts are usually metal-poor and star-forming dwarfs sharing some similarities with those harboring long gamma-ray bursts \citep[e.~g.][]{chenetal2013,chenetal2017,lunnanetal2014,leloudasetal2015,perleyetal2015,schulzeetal2018,schulzeetal2021,taggartetal2021,clelandetal2023}.

A fraction of SLSNe I show spectral features of helium. In some of them, the \ion{He}{i} $\lambda$10830 line has been identified in post-maximum near-IR (NIR) spectra, and such SNe were sometimes classified as SLSNe Ib\footnote{In the case of PTF10hgi, it has been also referred to as SLSN IIb due to the hydrogen and helium richness of its spectrum \citep{quimbyetal2018,galyam2019}.} \citep[see, e.g.,][]{quimbyetal2018,yanetal2020,kumaretal2025,kumaretal2026}. 
Interestingly, \citet{chenetal2023b} found that all the best densely sampled LCs of SLSNe Ib presented in \citet{yanetal2020} either showed LC rebrightenings\footnote{A similar association has also been seen in some normal stripped-envelope SNe \citep[see e.~g.][]{sollermanetal2020}.} \citep[see also ][]{zhuetal2023} or were better described by a hybrid $^{56}$Ni-decay + CSM interaction model. This suggests a link between He-line detections, bump occurrence, and CSM interaction in SLSNe. However, the current sample remains too small to draw firm conclusions about this. Therefore, it is crucial to expand and systematically study the sample of SLSNe Ib in order to constrain the nature of SLSN I(b) progenitors and their powering mechanisms.

We present and interpret the data of the observational campaign of the SLSN Ib SN~2021bnw, whose helium abundance was inferred from the analysis of a NIR spectrum observed 87 days after maximum. SN~2021bnw was located at $\mathrm{R. A.}=10^{\rm h}53^{\rm m}52^{\rm s}.18,\,\mathrm{Dec.}= +12\degr33\arcmin29\arcsec.12$, at a redshift $z=0.0888\pm0.0004$ (see Sec.~\ref{sec:spectroscopy}) and was discovered by \citet{fremlingetal2021} with the Zwicky Transient Facility \citep[ZTF;][]{grahametal2019,bellmetal2019a} and was classified as a hydrogen-poor SLSN I by \citet{mageeetal2021} within the advanced (extended) Public ESO Spectroscopic Survey of Transient Objects \citep[ePESSTO+, ][]{smarttetal2015}. SN~2021bnw is not grouped among the brightest SLSNe I as its absolute magnitude in the optical bands is $-20.7$ mag (see Sec.~\ref{sec:photometry}), and it was included by \citet{gomezetal2022} in a sample of luminous SNe, that is, SNe with an absolute peak magnitude between that of classical core-collapse SNe and the faintest SLSNe. In addition, \citet{pursiainenetal2023} and \citet{poidevinetal2023} included SN~2021bnw in two polarization studies, both of which suggested no significant departure from spherical symmetry for this event. Finally, \citet{neckeretal2022} initially included SN~2021bnw in a sample of possible counterparts of the high-energy neutrino IceCube \citep{aartsenetal2014,aartsenetal2015} Alert IC200109A \citep{garrappaetal2020}, but then it was not recognized as a viable candidate.

Hereafter, in Sec.~\ref{sec:photometry}, we discuss the observed and the bolometric LCs of SN~2021bnw. In Sec.~\ref{sec:spectroscopy} we present the optical and NIR spectra of SN~2021bnw. In Sec.~\ref{sec:discussion} we compare the spectro-photometric data of SN~2021bnw in the context of SLSNe I, and we discuss a possible model for its interpretation. We finally present our conclusion in Sec.~\ref{sec:conclusions}.
Throughout the paper, we assume a flat
Universe with $\Omega_{\rm M}= 0.31$ and $\Omega_{\Lambda}=0.69$ and with a Hubble constant $H_0=70.5\pm2.37\,\mathrm{km\,s^{-1}\,Mpc^{-1}}$ \citep{kethanetal2021}.
\section{Observed and bolometric light curves}
\label{sec:photometry}

\begin{figure*}
    \centering
    \includegraphics[width=1\textwidth]{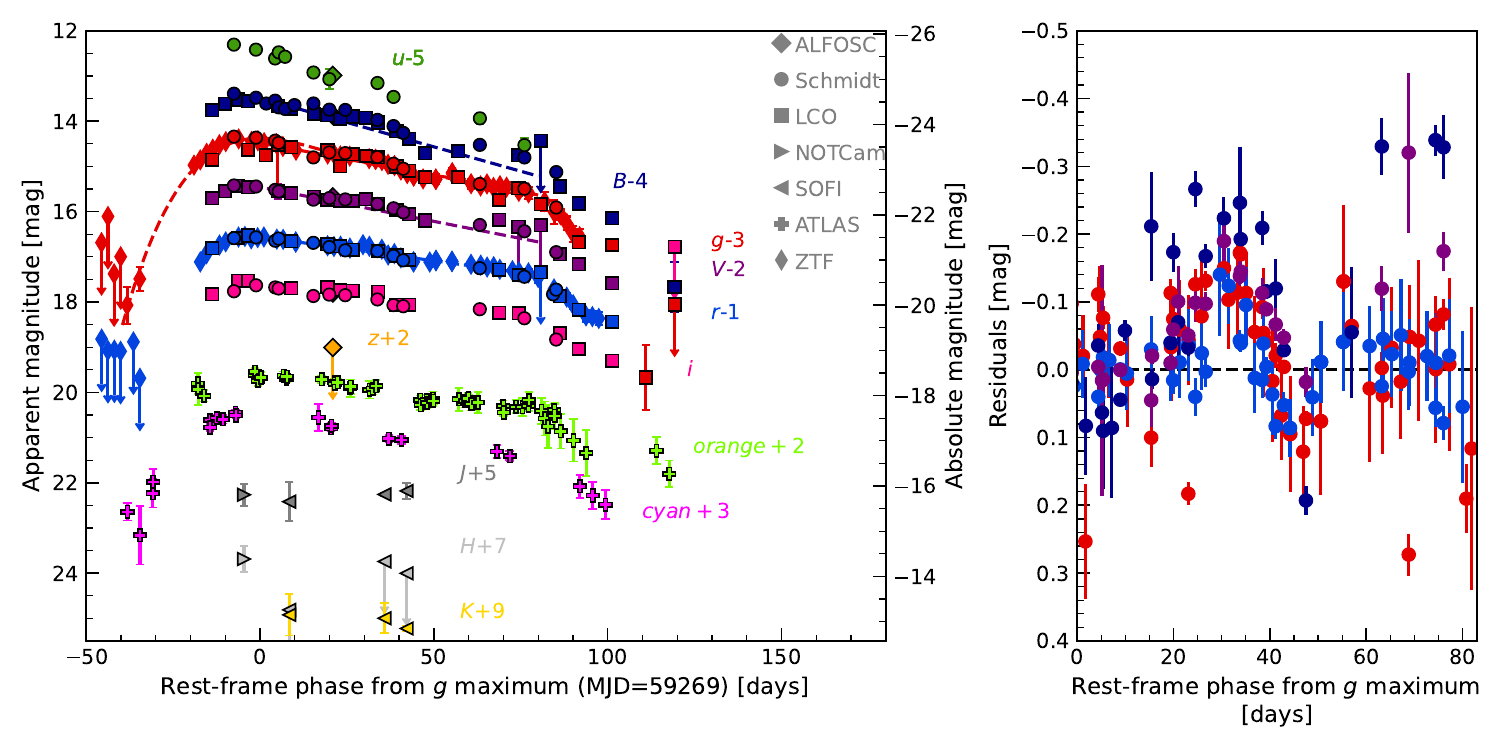}
    \caption{Left panel: $u$-, $B$-, $g$-, $V$-, $r$-, $i$-, ATLAS $cyan$-, ATLAS $orange$-, $z$-, $J$-, $H$-, $K$-filter LCs (green, dark blue, red, purple, blue, magenta, orange, light green, violet, dark gray, light gray, and yellow symbols) of SN~2021bnw. The LCs are displayed in apparent (left axis) and absolute magnitude (right axis) and are shifted by a constant (shown in the label). The dashed line shows the polynomial fit used to estimate the maximum luminosity epoch. The LCs in different filters are colored with different colors as labeled at the right side. The different symbols correspond to different instruments, as labeled in gray in the top right corner. The magnitudes are reported in the AB magnitude system. Right panel: Residuals after the subtraction of the best-fit one-dimensional polynomial between 2 and 80 days after maximum luminosity in $B,g,V,r$ bands. The dashed black line marks the zero-residual level for reference. In this panel, the dots are color-coded as in the left panel, and the detection limits were removed for clarity.}
    \label{fig:lcs_2021bnw}
\end{figure*}
\begin{figure}
    \centering
    \includegraphics[width=0.5\textwidth]{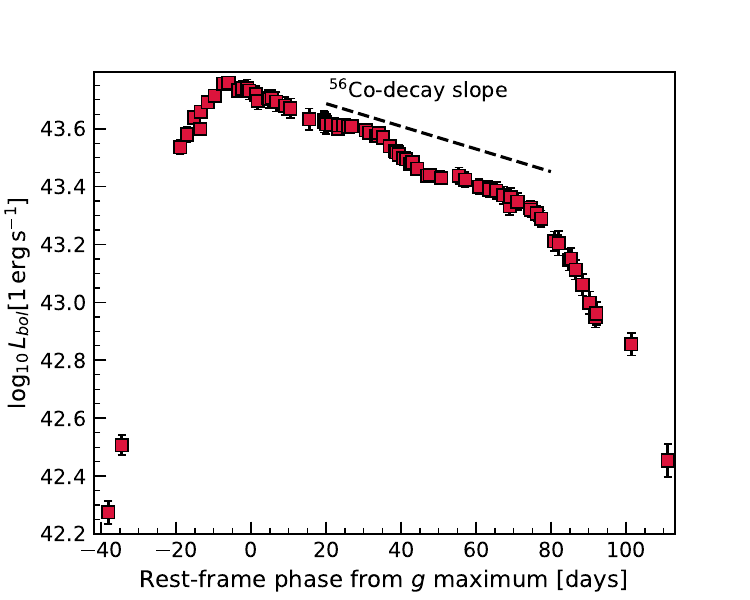}
    \caption{Pseudo-bolometric LC of SN~2021bnw. The $^{56}$Co slope is shown for comparison.}
    \label{fig:bolometric}
\end{figure}
We obtained the optical-to-NIR photometry of SN~2021bnw with an extensive photometric follow-up campaign using several facilities, and we reduced the imaging data with dedicated pipelines or the reduction package \texttt{ecsnoopy} (see Appendix~\ref{app:photo} for a detailed description). We estimated the epoch of maximum luminosity by performing a polynomial fit around the pre- and post-maximum epochs and found that the maximum luminosity of SN~2021bnw occurred in $g$ band on $\mathrm{MJD}=59269.12\pm0.01$ at $g=17.38\pm0.01$ mag. The errors on the peak magnitude and of the maximum-luminosity epoch were evaluated by varying the degree of the polynomial. After measuring the redshift by means of the narrow emission lines from the host galaxy (see Sec.~\ref{sec:spectroscopy}), we obtained the luminosity distance as $d_{\rm L}=402.5\pm17\,\mathrm{Mpc}$. As we identified no narrow absorption interstellar line of the Na ID doublet in the optical spectra, we assumed the host attenuation to be negligible \citep{poznanskietal2012}. Given the distance modulus $\mu=38.02\pm0.10$ mag and the Galactic absorption $A_{V}=0.05$ mag \citep{schlaflyetal2011}, we derived the color excess using a reddening law with $R_{\rm V}=3.1$ \citep{cardellietal1989,fitzpatrick1999} and computed an absolute peak $g$ magnitude of $M_g=-20.70\pm0.10$ mag. Moreover, we obtained the pseudo-bolometric LC in the following way: for each epoch and filter, we converted the multicolor photometry into flux, and we thus retrieved the spectral energy distribution (SED). We then integrated over the wavelength using the trapezoidal rule and assumed zero flux outside the integration boundaries. Finally, for each epoch, we obtained the pseudo-bolometric luminosity by multiplying the integrated flux by $4\pi d_{\rm L}^2$. The error bars on the pseudo-bolometric LC were estimated by combining the photometric error and the error on the luminosity distance. The pseudo-bolometric LC was computed adopting the epochs of the $g$-filter LC as a reference since it is the most densely sampled (Table~\ref{tab:opt_tables}), and missing coeval magnitude measurements in other filters were estimated by interpolation (or extrapolation, if necessary), assuming constant color. Our calculation of the pseudo-bolometric LC of SN~2021bnw did not include UV magnitudes. Because bright stars are lacking in the field, causing a spacecraft safety issue, it was not possible to trigger the \textit{Swift}/Ultra-violet Optical Telescope \citep[UVOT,][]{gehrelsetal2004} for SN~2021bnw.

The multiband LCs of SN~2021bnw are shown in Fig.~\ref{fig:lcs_2021bnw}. The pre-maximum detections are scarce, but enabled us to constrain the rest-frame rise time toward maximum luminosity. Given the last ZTF-$g$ detection limit at phase $\approx-40$ rest-frame days ($\mathrm{MJD}=59225.4$) before maximum and the first ZTF detection at phase $\approx-38$ rest-frame days ($\mathrm{MJD}=59227.4$) before maximum, we estimated a rise time for the $g$-filter LC of about $39$ days. Soon after maximum luminosity, the LCs show a linear declining phase lasting about 80 rest-frame days, during which the optical LCs presented two possible rebrightenings. In Fig.~\ref{fig:lcs_2021bnw} (right panel) we subtracted for each of the best-sampled LCs ($B,V,g,r$) the best-fit first-order polynomial between 5 and 80 days (excluding the epochs in which the bumps occur), and we recognized two deviations from the linear trend. The first deviation peaks at about 35 days, has an amplitude $\lesssim 0.25\,\mathrm{mag}$ in the $B$ filter, and slightly decreases in the redder filters. The second deviation is seen in $B$ and $V$ filters only around $70-80$ days, but with a slightly larger amplitude of $\lesssim 0.3\,\mathrm{mag}$ in $B$ band and a comparable one in $V$ band (when we neglect the two $V$ detections with higher error bars; see the right panel in Fig.~\ref{fig:lcs_2021bnw}). After this phase, the LCs steepened for about 30 days, after which the SN set behind the Sun, and thereafter, it faded beyond the detection limit. The pseudo-bolometric LC of SN~2021bnw is shown in Fig.~\ref{fig:bolometric}. At about maximum luminosity, it exhibited a wedge-shaped rise and decline, which is not new among SLSNe I (see Sect.~\ref{sec:comparisons}), and it reproduced the behavior of the multiband LCs in $g,r$ filters overall. The average slope of the post-maximum declining phase within 80 days of the bolometric LC of $\approx0.016\,\mathrm{mag\,d^{-1}}$ is faster than the $^{56}\mathrm{Co}$-decay slope assuming full $\gamma$-ray trapping. During this phase, the pseudo-bolometric LC slightly oscillated at $\approx40$ days, but the second bump at $\sim75$ days (see Fig.~\ref{fig:bolometric}) is not evident or might be blended with the declining knee, after which the LC declined with a slope of $\approx0.07\,\mathrm{mag\,d^{-1}}$.

\section{Optical and NIR spectra}
\label{sec:spectroscopy}
\begin{figure*}
    \centering
    \includegraphics[width=1\textwidth]{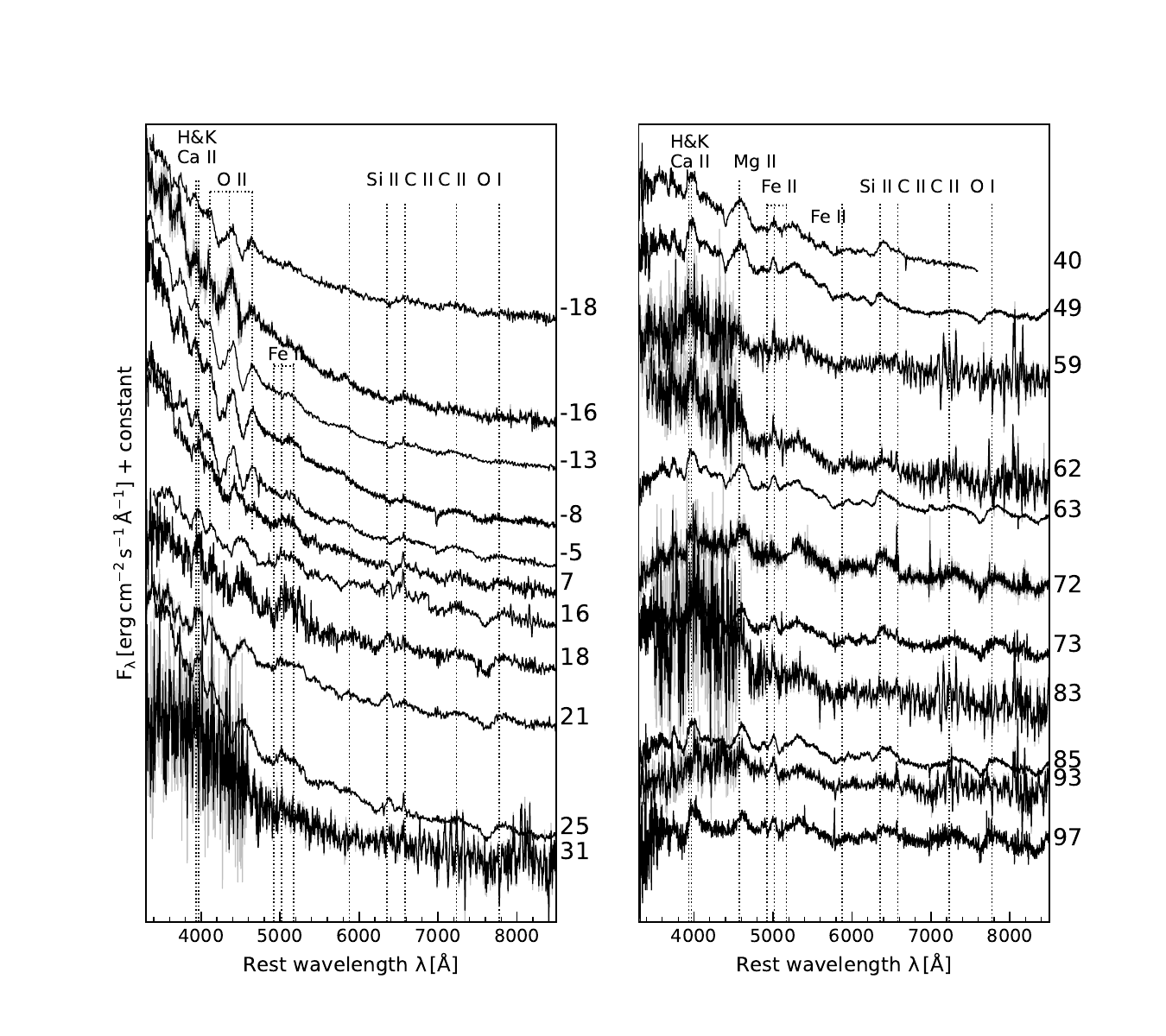}
    \caption{Spectral evolution of SN~2021bnw. The dotted lines mark the line identifications labeled at their top and refer to the emission component at rest frame. On the right side of each spectrum, we report the rest-frame phase with respect to the $g$-band maximum luminosity (in days). The spectra at phases $-12$, 25, 35, 62, 66, 72, 86, and 96 were rebinned (with a bin size of 5 \AA{}) and smoothed with a Savitzky-Golay filter for clarity. Left panel: Spectra from -18 days to 31 days. Right panel: Spectra from 40 to 97 days.}
    \label{fig:specs_2021bnw}
\end{figure*}
We obtained optical spectroscopy of SN~2021bnw with an extensive photometric follow-up campaign using different facilities, and we reduced the raw spectroscopic frames with standard \texttt{IRAF} routines or dedicated pipelines (see Appendix~\ref{app:spectra} for a detailed description).
The spectral evolution of SN~2021bnw is shown in Fig.~\ref{fig:specs_2021bnw}. The hot ($\sim14,000-17,000$ K) pre-maximum/maximum spectra show the \ion{O}{ii} and the \ion{Ca}{ii H\&K} features on the blue side and broad \ion{C}{ii} $\lambda\lambda6580, 7121$, \ion{O}{i} $\lambda7774$ lines on the red side. No \ion{O}{ii} lines are seen in the almost featureless spectrum observed 7 days after maximum, although the lower signal-to-noise ratio of the blue side does not rule out their contribution. On day 16, the \ion{Si}{ii} $\lambda6355$ line also appears, and on days 25 and 31, the spectral continuum suddenly becomes bluer than on day 21, approximately coeval with the first LC bump. It is not clear whether this also occurred in the spectra observed at $70-80$ days, when the second bump appeared in the $B,V$ LCs, or if this is an artifact of the lower signal-to-noise ratio on the blue side of the spectrum. As noted in the case of SN~2017gci \citep{fioreetal2021} showing two bumps in the post-maximum LC, no noticeable changes are seen in the spectra during the epochs in which the bumps appear. Then, on day 31, the spectrum lacked any discernible features and gradually transitioned into a new phase mirroring the spectrum of an SN Ic at maximum luminosity. This is shown by the emergence of the \ion{Mg}{ii} at $\sim4570\,\text{\AA{}}$ and \ion{Fe}{ii} features between $4500-5500\,\text{\AA{}}$. In particular, in the spectrum 40 days after maximum, a broad feature at $\approx5010\,\text{\AA{}}$ appeared, which we tentatively interpret as \ion{Fe}{ii} $\lambda\,5018$, although the other lines of the triplet ($\lambda\lambda\lambda\,4923,5018,5169$) are not seen or might be blended. These features are seen until day 97, when the spectrum was still photospheric, while the continuum gradually cooled and the \ion{O}{i} $\lambda\,7774$ became more pronounced. During the latest phases, the narrow emission lines from the host galaxy start to be seen. We measured the redshift of the host galaxy of SN~2021bnw using the narrow H$\alpha$ and obtained $z=0.0888\pm0.0004$ by fitting a Gaussian to the line profile.
\begin{figure}
    \centering
    \includegraphics[width=0.5\textwidth]{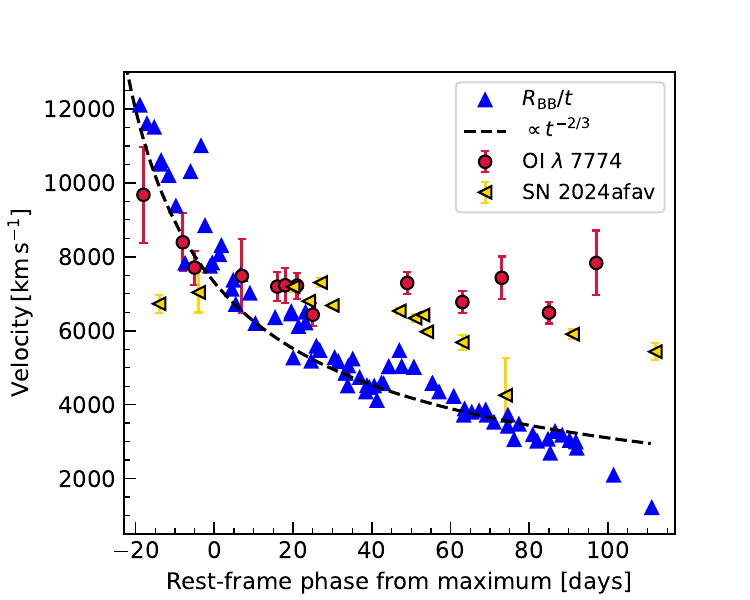}
    \caption{Velocities of SN~2021bnw measured by fitting a Gaussian to the absorption minima of the \ion{O}{i} $\lambda\,7774$ (red dots) and from the blackbody radius $R_{\rm BB}$(blue triangles; see Fig.~\ref{fig:photosphere}). For comparison, we also plot the \ion{O}{i} $\lambda\,7774$ velocity of SN 2024afav \citep[gold triangles;][]{kumaretal2026} and a $\propto t^{-2/3}$ curve (dashed black line).}
    \label{fig:vphot2}
\end{figure}
We used the absorption minima of \ion{O}{i} P-Cygni profiles to estimate the \ion{O}{i} velocity of SN~2021bnw (see Fig.~\ref{fig:vphot2}). In detail, we fit a Gaussian to the absorption profiles and measured the shift compared to the rest-frame emission wavelength.
In addition, we estimated the expansion velocity from the blackbody radius (see Sec.~\ref{sec:photosphere}), which approximately follows a power law with an index $-2/3$ (see Fig.~\ref{fig:vphot2}).
The two velocities agree within $\sim20$ days after maximum and are consistent with that of the SLSN Ib SN 2024afav (see also Sec.~\ref{sec:comparisons}). After this, the \ion{O}{i} velocity remained nearly constant at $\sim7500\,\mathrm{km\,s^{-1}}$, while the blackbody velocity still followed the power-law behavior. From this comparison, we broadly constrained the initial photospheric velocity to $\sim10,000-7,000\,\mathrm{km\,s^{-1}}$ at pre-maximum/maximum epochs.
\begin{figure*}
    \centering
    \includegraphics[width=1\textwidth]{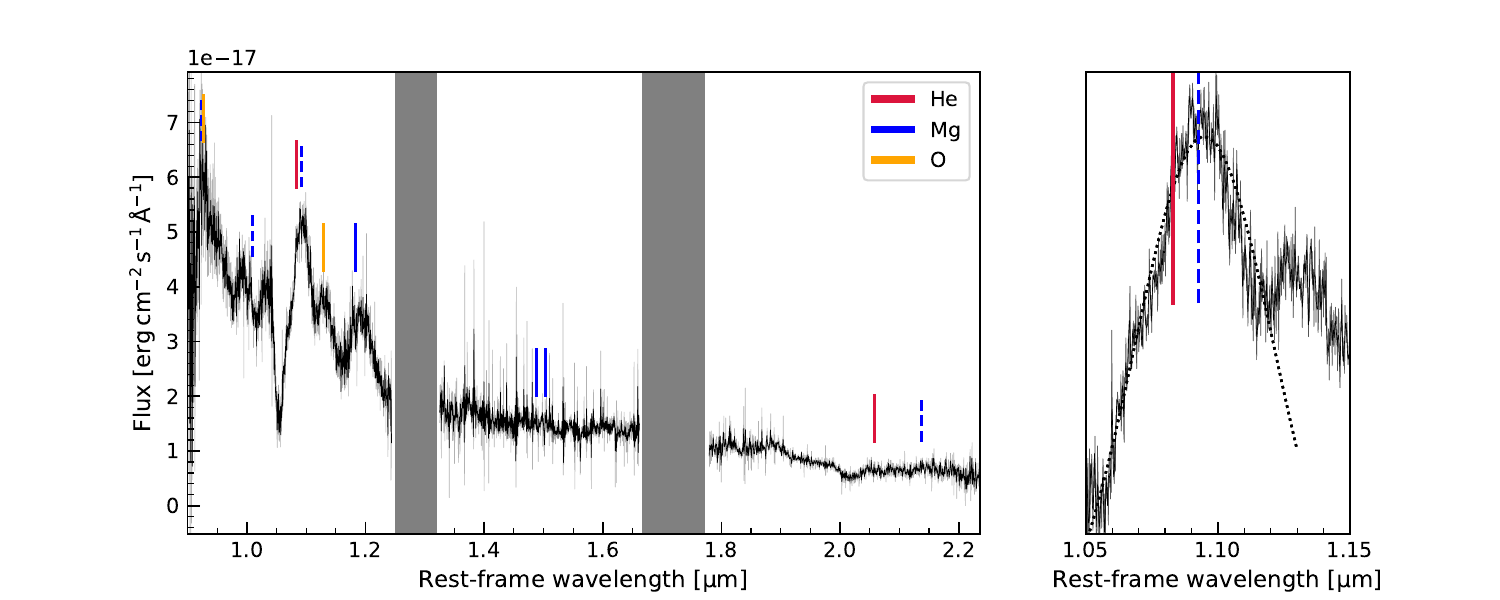}
    \caption{Left panel: NIR spectrum of SN~2021bnw 87 days after maximum. The gray shaded areas mark the regions contaminated by telluric absorptions. Right panel: Same as the left panel but zooming into the region close to 1.1 $\mu$m. In both panels, the wavelengths corresponding to the emission components of the NIR features from helium, magnesium, and oxygen are marked with different colors (see the legend in the top right corner), and the solid and dashed lines correspond to neutral and singly ionized features, respectively.}
    \label{fig:nirspectrum}
\end{figure*}
The NIR spectrum (see Fig.~\ref{fig:nirspectrum}, left panel) shows five absorptions between $0.9-1.3\,\mathrm{\mu m}$ that are usually seen in the NIR spectra of stripped-envelope SNe \citep[e.g.,][]{shahbandehetal2022}. However, given the high expansion velocities, some features are likely blended, and their identification can be challenging. This is true in particular for the prominent P-Cygni feature at $\sim 1.1\,\mathrm{\mu m}$. A Gaussian fit of its emission component is peaked at $\lambda\simeq1.095\,\mathrm{\mu m}$ and indicates a contribution of \ion{Mg}{ii} $\lambda\,1.0927\,\mathrm{\mu m}$ (see Fig.~\ref{fig:nirspectrum}, right panel).
A closer inspection of the blue wings of the emission (Fig.~\ref{fig:nirspectrum}, right panel) draws attention to a possible inflection about $\sim1.08\,\mu\mathrm{m}$, likely attributable to the contribution of the \ion{He}{i} $\lambda\,1.083\,\mu\mathrm{m}$ P-Cygni profile. Moreover, in the red wing of the emission component at $~1.1\,\mathrm{\mu m}$, we tentatively identify the weak bump as \ion{O}{i} $\lambda\,1.1290\,\mathrm{\mu m}$ and the following absorption as \ion{Mg}{i} $\lambda\,1.1828\,\mathrm{\mu m}$, while blueward of it, we identify the contribution of \ion{O}{i} $\lambda\,0.9264\,\mathrm{\mu m}$+\ion{Mg}{ii} $\lambda\,0.9227\,\mathrm{\mu m}$ and \ion{Mg}{ii} $\lambda\,1.0092\,\mathrm{\mu m}$. Finally, the identification of the \ion{He}{i} line is corroborated by the weaker absorption at $\sim2.02\,\mathrm{\mu m}$, which is compatible with \ion{He}{i} $\lambda\,2.0581\,\mathrm{\mu m}$ (see also Sec.~\ref{sec:spectroscopy}). However, optical \ion{He}{i} lines such as the $\lambda\,5876$ line usually seen in SNe Ib are not clearly distinguished in the spectra of SN~2021bnw, but they are likely blended with iron/magnesium features (see also Sec.~\ref{sec:spectroscopy}). In any case, the relative weakness of the \ion{He}{i} $\lambda\,5876$ feature does not necessarily invalidate the \ion{He}{i} identification in the NIR spectrum \citep[see, e.g.,][]{teffsetal2020}.

To test the proposed line identifications in the NIR spectrum of SN~2021bnw, we computed a synthetic spectrum using the Monte Carlo radiative-transfer code \texttt{TARDIS} \citep{kerzendorfandsim2014}. The open-source one-dimensional code \texttt{TARDIS} is widely used for the rapid spectral synthesis of supernovae. Despite some simplifying assumptions such as spherical symmetry and homologous expansion, \texttt{TARDIS} is capable of consistently describing the plasma state and handling radiation-matter interactions. It defines a computational domain in velocity space with a fixed inner boundary, representing the SN photosphere. From this boundary, a blackbody radiation field propagates and interacts with the overlying ejecta.

In detail, we used the \texttt{TARDIS} built-in power-law density profile $\rho(v,t)=\rho_0(t/t_0)^{-3}(v/v_0)^{-\alpha}$ to calculate the spectrum.
 We inferred $\alpha=4$ from the velocity evolution (see Sec.~\ref{sec:spectroscopy}) using a formalism similar to that by \citet[][see Eq. 4 therein]{nicholl2026}, $v_0=10\,000\,\mathrm{km\,s^{-1}}$, and fixed $\rho_0=3.4\times10^{-9}\,\mathrm{g\,cm^{-3}}$. The last two parameters were varied to obtain a \texttt{TARDIS} ejecta mass lower than or similar to that reported in Sec.~\ref{sec:scenario}. The other \texttt{TARDIS} simulation parameters were set up following \citet{williamsonetal2021} and \citet{boyleetal2017}, and they are summarized in Table~\ref{tab:tardis}. We adopted a homogeneous abundance profile and the latest version (March 2025) of the \texttt{TARDIS} atomic-data file \citep{kurucz2017}. In addition to helium, we input the chemical elements that cause the lines identified in the optical spectra and varied their abundances to fit the NIR spectrum of SN~2021bnw. To do this, we also used the optical spectrum at 85 days as an additional reference and explored the \texttt{TARDIS} parameter space to provide a good compromise between the optical and NIR spectral regions. Good spectral fits were obtained with a helium mass of $\sim 0.10-0.15\,\mathrm{M_\odot}$ by varying the outer \texttt{TARDIS} integration boundary. However, while our He-mass estimate might be quite uncertain due to the simplifying \texttt{TARDIS} assumptions described above, it is comparable to the threshold of non-hidden helium estimated by \citet{teffsetal2020} for optical He features to become visible in the spectra of stripped-envelope SNe, and it is about two/three times the upper limit on the He mass found by \citet{kumaretal2025b} from the He-devoid NIR spectrum of the SLSN I SN 2024ahr.
 We emphasize that we did not attempt to model the entire spectrum, but only tested the helium line  identification.
\begin{figure*}
    \centering
    \includegraphics[width=1\textwidth]{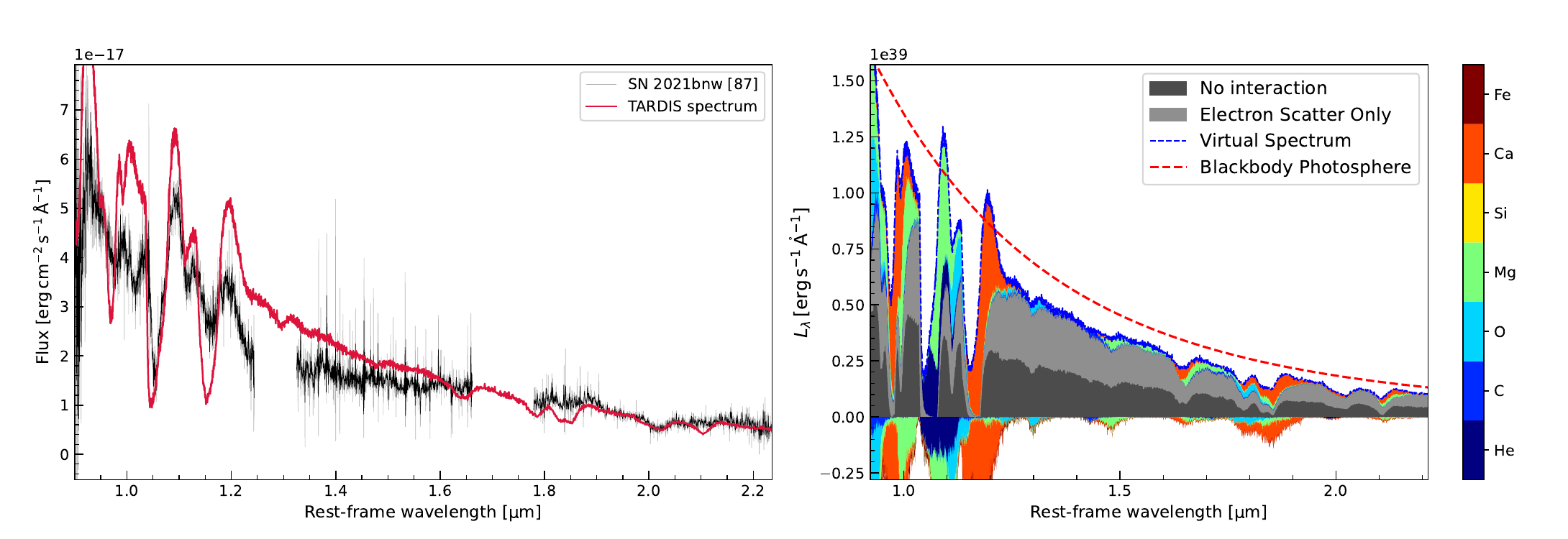}
    \caption{Left panel: Comparison of the NIR spectrum of SN~2021bnw at 87 days after maximum with the \texttt{TARDIS} synthetic spectrum. Right panel: Spectral decomposition plot \citep{kromeretal2013} where different contributions to
emission and absorption of quanta in the \texttt{TARDIS} Monte Carlo simulation are colored as in the palette. }
    \label{fig:SDECplot}
\end{figure*}
\begin{table}[]
   \centering
   \caption{\texttt{TARDIS} configuration used for the synthetic spectrum in Fig.~\ref{fig:SDECplot}.}
   \begin{tabular}{ll}
       \hline        \hline
        luminosity requested &  $1.65\times10^{43}\,\mathrm{erg\,s^{-1}}$\\
        time\_explosion & 126 days\\
        $\rho_0$& $3.4\times10^{-9}$ $\mathrm{g\,cm^{-3}}$\\
        $v_0$&$10\,000\,\mathrm{km\,s^{-1}}$\\
        $t_0$&1 day\\
        exponent & -4\\
        ionization & nebular \\
        excitation & dilute-lte \\
        radiative rates type & dilute-blackbody \\
        line interaction type & downbranch \\
        helium treatment & recomb-nlte \\
        \hline         \hline
   \end{tabular}
   
   \label{tab:tardis}
\end{table}

The observed NIR spectrum is compared with the synthetic \texttt{TARDIS} spectrum in Fig.~\ref{fig:SDECplot}. The \texttt{TARDIS} spectrum predicts the contribution of \ion{He}{i} to the absorption at 1.05 $\mu$m, although it partly overlaps with \ion{Mg}{ii}. By contrast, helium alone reproduces the absorption at 2 $\mu$m reasonably well (see Fig.~\ref{fig:SDECplot}, right panel). Since this region is expected to be very little contaminated by other features \citep{shahbandehetal2022}, we can more confidently attribute contributions in the NIR spectrum of SN 2021bnw to helium.
\section{Discussion}
\label{sec:discussion}
\subsection{Photometric and spectroscopic comparisons with other SNe}
\label{sec:comparisons}
\begin{figure}
    \centering
    \includegraphics[width=0.5\textwidth]{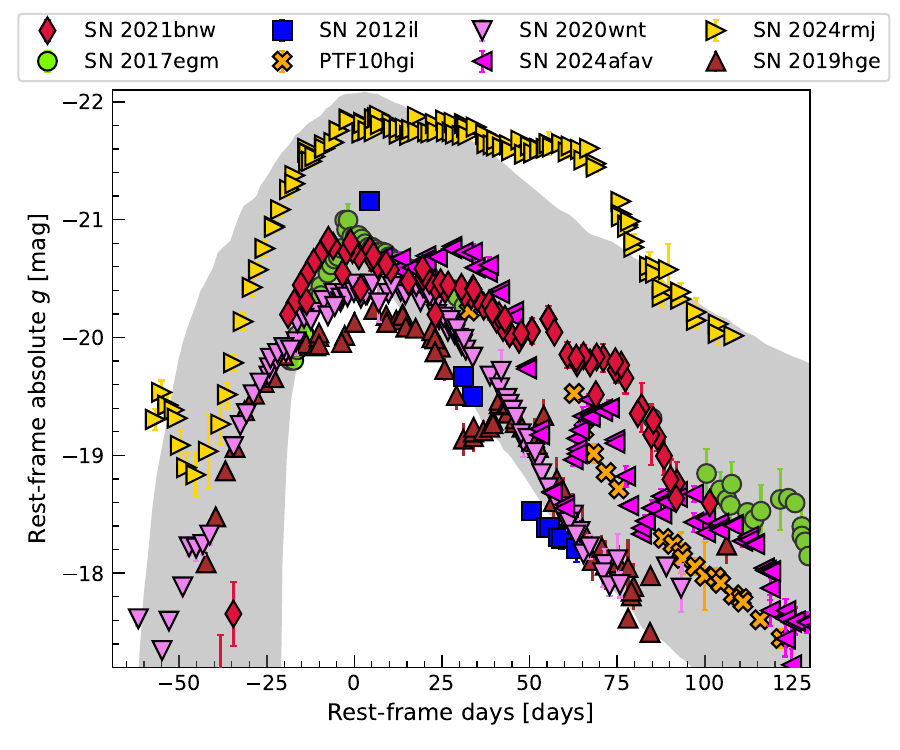}
    \caption{Comparisons of the $K$-corrected $g$-band absolute LCs of SN~2021bnw with those of the comparison sample. The LCs of the selected SNe are marked differently, as explained in the caption at the top. The gray region denotes the range encompassed by the $g$ mean absolute LC of the SLSN-I sample analyzed by \citet{gomezetal2024}}.
    \label{fig:phot_comparisons}
\end{figure}
 \begin{figure}
     \centering
     \includegraphics[width=0.9\linewidth]{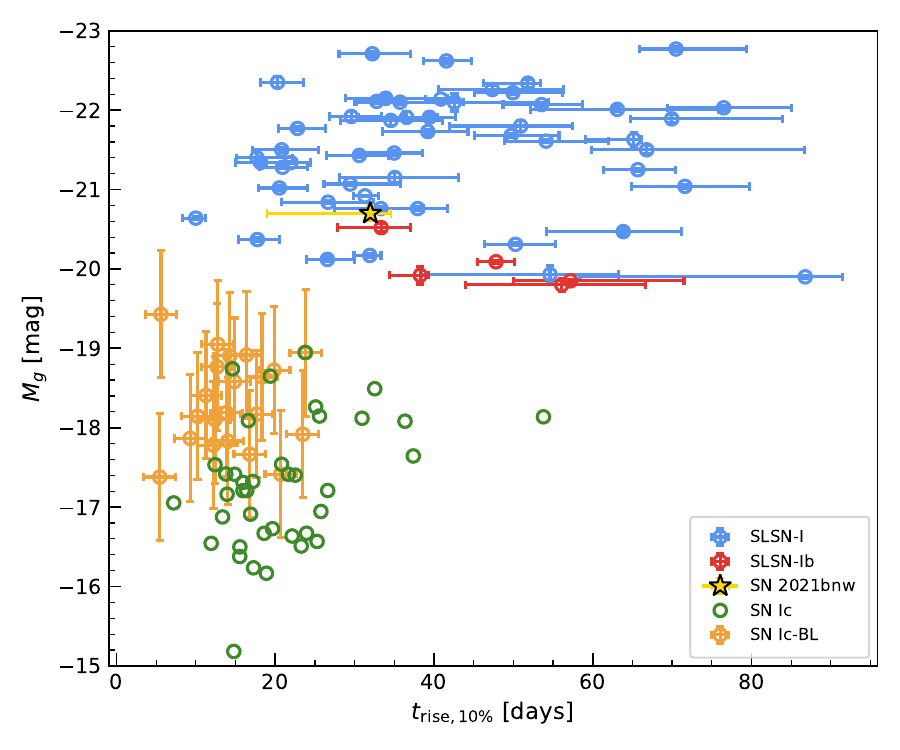}
     \caption{$t_{{\rm rise}, 10\%}$ vs. $M_g$ ($g$-band), adapted from \citet{chenetal2023a}. SN~2021bnw is shown as a yellow star. SLSNe~I/Ib (light blue/red circles) are plotted alongside SNe~Ic/Ic~BL from \citet{taddiaetal2019} and \citet{barbarinoetal2021}.}
     \label{fig:chen_et_al}
 \end{figure}

\begin{figure}
    \centering
    \includegraphics[width=0.4\textwidth]{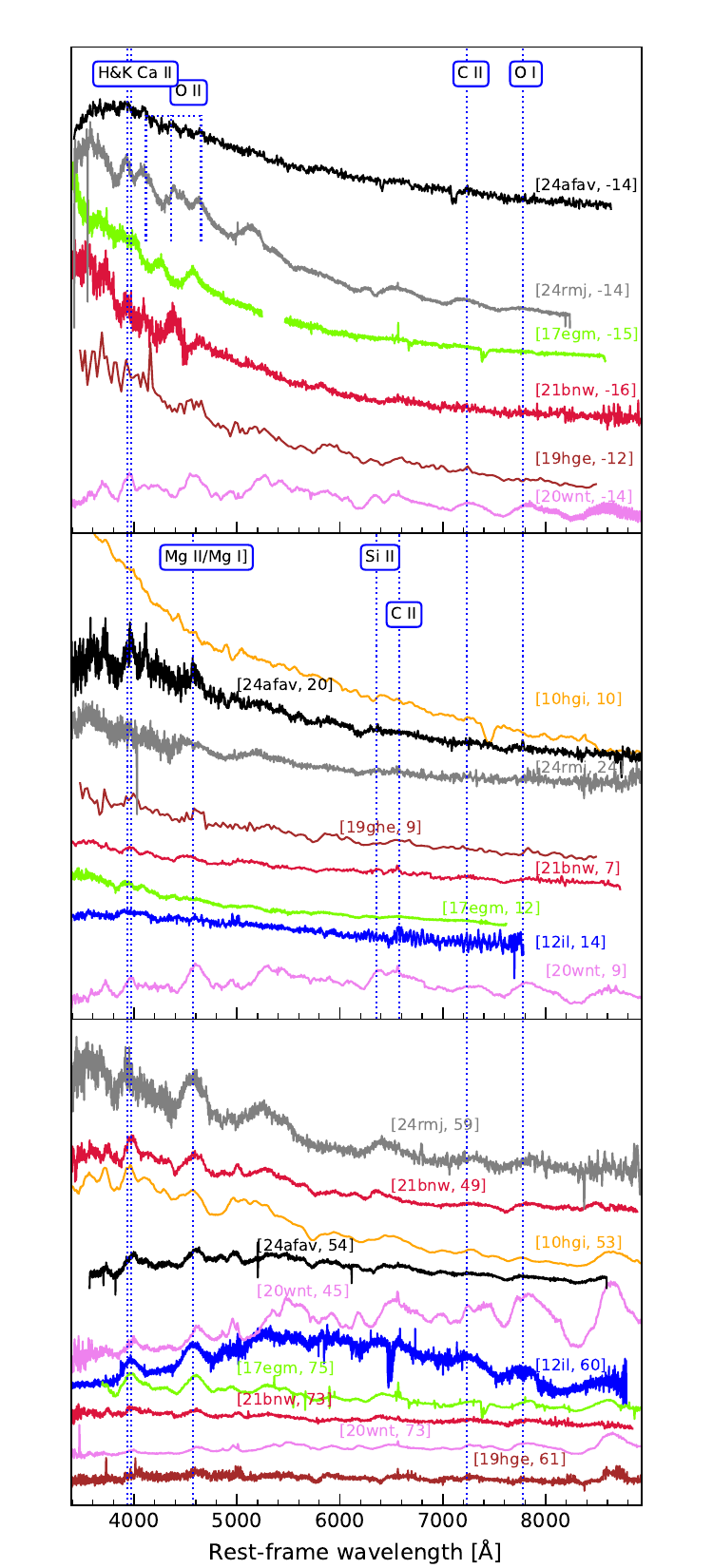}
   \caption{Comparison of optical spectra of SN~2021bnw with those of the comparison sample (see text for details) at different phases: pre-maximum (top panel), post-maximum (within 15 days after maximum, middle panel), and post-maximum (within 80 days after maximum, bottom panel). The name and rest-frame phase from maximum of each spectra is labeled between square brackets next to each spectrum. The rest-frame wavelength of some transitions is marked with dotted blue lines.}
    \label{fig:spectral_comparisons_opt}
\end{figure}

\begin{figure}
    \centering
    \includegraphics[width=0.5\textwidth]{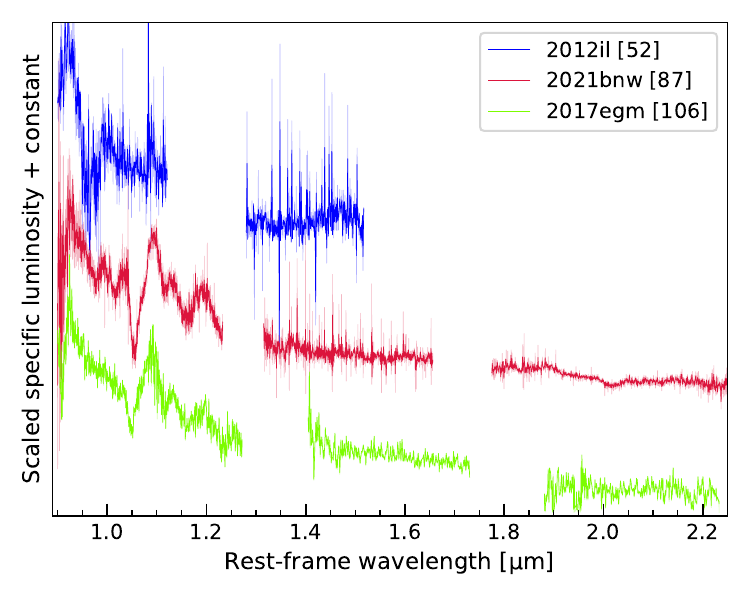}
    \caption{Same as Fig.~\ref{fig:spectral_comparisons_opt} but for NIR spectra. The spectra of SN~2012il and SN~2021bnw are smoothed with a Savitzky-Golay filter. The spectrum of SN 2021il was also rebinned for clarity because its signal-to-noise ratio is low.}
    \label{fig:spectral_comparisons_NIR}
\end{figure}
We selected a sample of SLSNe Ib to compare the photometry and spectra of SN~2021bnw with those of analogous objects with post-maximum He-rich NIR spectra: PTF10hgi, SN 2012il \citep{inserraetal2013}, SN 2017egm \citep{nicholletal2017a,boseetal2018,hosseinzadehetal2022,zhuetal2023}, SN~2019hge \citep{yanetal2020}, SN 2024rmj \citep{kumaretal2025}, and SN 2024afav \citep{kumaretal2026}. We also added the SLSN I SN~2020wnt \citep{gutierrezetal2022} to the comparison sample, which has possible \ion{He}{I} detections at about $\sim5000$ and $\sim7000\,\text{\AA{}}$ and a similar steepening in the LC.
We compared the $K$-corrected absolute LCs in all available filters for the SNe of the comparison sample (Fig.~\ref{fig:phot_comparisons}). For SN~2021bnw, we used the $K$-correction following \citet[see our Table~\ref{tab:speclog}]{fioreetal2021}, while for the other SNe, we used the approximation $K=-2.5\log_{10}(1+z)$ \citep{hoggetal2002}, which is reasonable for SLSNe I \citep[][]{chenetal2023a}. Interestingly, the LCs of SN~2021bnw and SN~2017egm are remarkably similar in terms of absolute magnitude around the pre-maximum/maximum and early post-maximum epochs ($\lesssim20$ days), in which they present a wedge shape that is sharper in $g$ and broader in other filters. However, the LC of SN~2017egm rises faster toward maximum than that of SN~2021bnw (see Fig.~\ref{fig:phot_comparisons}). After $\sim20$ days, the follow-up of SN~2017egm was interrupted, but soon after $\sim85$ days, their $g$ LCs were still similar until $\sim100$ days. Compared to them, the LCs of SN~2012il, PTF10hgi, and SN~2020wnt evolved faster after maximum, while the LC of SN 2024rmj, the brightest of the sample, reached a plateau-like phase after maximum. Moreover, the LCs of SN 2019hge and SN 2024afav exhibited sharp bumps after maximum.

To further characterize SN~2021bnw, we reproduced the analysis of \citet{chenetal2023a} by measuring the time intervals corresponding to fractional flux levels of $x=0.1$ and $x=e^{-1}$ relative to the peak absolute $g$ flux during its rising and declining phases ($t_{{\rm rise}, x}$ and $t_{{\rm decline}, x}$). We fitted a fifth-order polynomial to the entire LC and assigned conservative error bars to $t_{\rm rise}$, equal to the time interval between the estimated $t_{\rm rise}$ and the preceding or subsequent detections because data before maximum are scarce. Next, we reproduced Fig.~6 of \citet{chenetal2023a} by plotting the $K$-corrected peak magnitude, $M_g$, against $t_{{\rm rise}, 10\%}$ (Fig.~\ref{fig:chen_et_al}). We note that the position of SN~2021bnw ($t_{{\rm rise},10\%}=31.98^{+3}_{-13}$ days, $M_g=-20.65\pm0.1$ mag) falls well within the region populated by other SLSNe Ib. Furthermore, while it follows the $t_{{\rm rise},1/e}$ versus\ $t_{{\rm rise},10\%}$ correlation within $0.24\sigma$ ($0.68$ days, assuming $1\sigma=2.84$ days as in \citealt{chenetal2023a}), SN~2021bnw is a clear outlier in the $t_{{\rm decline},1/e}$ versus\ $t_{{\rm rise},1/e}$ parameter space due to its slow post-maximum decline of $\sim0.016\,\mathrm{mag\,d^{-1}}$ (see Sec.~\ref{sec:photometry} and further discussion in Sec.~\ref{sec:scenario}).

 We also compared four representative spectra of SN~2021bnw with those of the SNe of the comparison sample in comparable phases (see Fig.~\ref{fig:spectral_comparisons_opt}, ~\ref{fig:spectral_comparisons_NIR}) after correcting them for redshift and Galactic extinction following \citet{schlaflyetal2011}. The spectra were obtained via \texttt{Wiserep}\footnote{\texttt{https://www.wiserep.org}.} \citep{yaronandgalyam2012}. The spectra of SN~2021bnw look more similar in terms of features and spectral continuum to those of SN~2019hge, SN~2017egm, and SN~2012il. However, SN~2012il evolved faster than SN~2021bnw and reached the featureless phase before SN~2021bnw (see Fig.~\ref{fig:specs_2021bnw}). The spectra of SN~2020wnt are redder and with stronger features in post-maximum epochs. Finally, we compared the NIR spectrum of SN~2021bnw with two spectra of SN~2017egm and SN~2012il, although in different rest-frame phases with respect to maximum. In the three spectra (see Fig.~\ref{fig:spectral_comparisons_NIR}), a feature is seen at about $\sim1.1\,\mathrm{\mu m}$, although with different characteristics. While it looks like a pure emission feature for SN 2012il, a P-Cygni profile is present for SN 2021bnw and SN 2017egm with a blue wing broader than the red one. The \ion{O}{i} $\lambda\,0.9264\,\mathrm{\mu m}$+\ion{Mg}{ii} $\lambda\,0.9227\,\mathrm{\mu m}$ feature is probably present in the NIR spectra of SN~2012il and SN~2017egm, while the \ion{O}{i} and \ion{Mg}{ii} features are identified in that of SN~2017egm alone because this wavelength range is not covered by the spectrum of SN~2012il. However, due to the low signal-to-noise ratio of the spectrum of SN~2012il and the considerable phase shift with respect to maximum, we cannot exclude that these SNe shared more or fewer spectral properties at coeval epochs.
\subsection{Blackbody temperature and radius}
\label{sec:photosphere}

\begin{figure}
    \centering
    \includegraphics[width=0.4\textwidth]{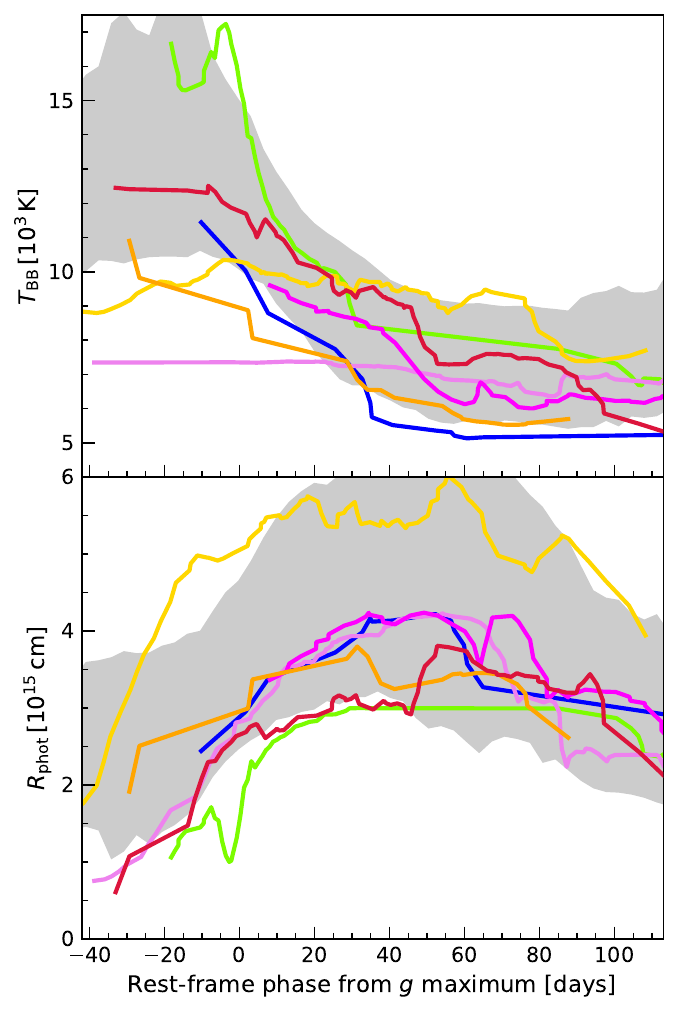}
    \caption{Evolution of the blackbody temperature of SN~2021bnw $T_{\rm BB}$ (top) and the blackbody radius of SN~2021bnw $R_{\rm phot}$ (bottom). Both quantities were smoothed with a Savitzki-Golay filter. The color convention follows that of Fig.~\ref{fig:phot_comparisons}. In each panel, the gray regions have the same meaning as in Fig.~\ref{fig:phot_comparisons}. }
    \label{fig:photosphere}
\end{figure}

We compared the blackbody radius and temperature of SN~2021bnw with those of the other SNe of the sample we considered and with those of the SLSN-I sample of \citet{gomezetal2024}.
For consistency, we used the same script to integrate the multiband photometry and retrieve the blackbody temperature and radius adopted for SN~2021bnw. 
We computed the evolution of the blackbody temperature and radius by fitting a Planck law to the SED (see Fig.~\ref{fig:photosphere}, top and bottom panel, respectively) and using the Stefan-Boltzmann law.

The early measurements of the blackbody temperature and radius of SN~2021bnw agree well with those of SN~2017egm. The apparent plateau behavior before maximum of SN~2021bnw is not reliable. These epochs are not even very well sampled except in the $g$ and $cyan$ filters (see Fig.~\ref{fig:lcs_2021bnw}), where the SED was extrapolated assuming constant color when coeval magnitude measurements in the other bands were lacking. The same is true for SN~2012il. After maximum, the blackbody temperatures of all the SNe smoothly declined at comparable rates toward an almost constant plateau. This 7,000-8,000 K temperature floor \citep{inserraetal2013,nicholletal2017} is quite common among SLSNe I and has been ascribed to oxygen recombination. This interpretation might be questioned for SN~2020wnt, as it also presents a temperature floor, but shows no prominent W-shaped \ion{O}{ii} features around its maximum-luminosity epoch. Similarities are also seen in the evolution of the blackbody radius (see Fig.~\ref{fig:photosphere}, bottom panel), which, in the case of SN~2021bnw, culminates its expansion phase about three days after maximum luminosity, and like the temperature, broadly agrees with the blackbody radii of SN~2017egm and SN~2012il up to $\sim40$ days after maximum. We caution that the measurements of blackbody temperatures and radius for SN~2021bnw might be inaccurate due to the lack of UV photometry. However, the blackbody radius and temperature of SN~2021bnw agree reasonably well with those of the SLSNe I analyzed by \citet{gomezetal2024}.
\subsection{Exploring a scenario for SN~2021bnw}
\label{sec:scenario}
Several properties deduced for SN~2021bnw in Secs.~\ref{sec:photometry}, \ref{sec:spectroscopy}, and \ref{sec:photosphere}, such as luminosity, spectral evolution, blackbody radius, temperature, and photospheric velocity, are consistent with those of the bulk SLSN-I population (Figs.~\ref{fig:phot_comparisons}, \ref{fig:photosphere}, \ref{fig:spectral_comparisons_opt}, \ref{fig:spectral_comparisons_NIR}; see also the photospheric velocities presented in \citealt{aameretal2025}). However, SN~2021bnw deviates from the $t_{{\rm decline},1/e}$ vs.\ $t_{{\rm rise},1/e}$ correlation, similar to four out of five SLSNe Ib analyzed by \citet{chenetal2023a}. \citet{chenetal2023a} suggested that the LCs of these SLSNe Ib might be shaped by interaction with a He-rich CSM, thus departing from a more standard LC shape that produces this correlation. Following this idea, we considered the combination of a central heating source and CSM interaction\footnote{This scenario was also used to model its close sibling SN~2017egm (see Sec.~\ref{sec:comparisons}).}. This hypothesis is supported in both cases by the sudden decline $\sim80$ days after maximum, which might mark the breakout of the forward shock from the CSM \citep[e.~g.][]{chevalier1982,chatzopoulosetal2012}. 

Considering a $^{56}$Ni+CSM scenario, \citet{wheeler2017egm} found for SN~2017egm a $^{56}$Ni mass $M(^{56}{\rm Ni})=0.15\,{\rm M}_\odot$, an ejecta mass $M_{\rm ejecta}=10.7\,{\rm M}_\odot$, and a CSM mass $M_{\rm CSM}=2.7\,{\rm M}_\odot$ using the module \texttt{csmrads0} of \texttt{TigerFit} \citep{chatzopoulosetal2013}. This code computes synthetic SN LCs exploiting the diffusion scheme introduced by \citet{arnett1980,arnett1982} for a variety of power mechanisms. Similarly, \citet{zhuetal2023} used the same tool, but on a more extended pseudo-bolometric LC, and found $M(^{56}{\rm Ni})=1\,{\rm M}_\odot$, $M_{\rm ejecta}=30\,{\rm M}_\odot$, and $M_{\rm CSM}=0.8\,{\rm M}_\odot$. We did not perform a fit with \texttt{csmrads0} as the LCs of SN~2017egm and SN~2021bnw almost overlap around maximum (Fig.~\ref{fig:phot_comparisons}), and we expect similar parameter estimates. However, although semi-analytic diffusion schemes can capture the general features of the observed LCs, their inferred physical quantities may substantially differ from those obtained with detailed radiative-transfer calculations \citep[see e.g.,][]{2013moriya}. \citet{kozyrevaetal2026} presented radiative-transfer calculations with the code \texttt{STELLA} \citep{blinnikovetal1998,blinnikovetal2000} to improve the reliability of the predictions on the physical parameters of the explosion. They also discuss in more detail the model and a systematic exploration of the parameter space for SN~2021bnw. In detail, their calculations converged on an ejecta mass of $\approx15-20$ \Msun{}, a CSM mass of $\approx7$ \Msun{}, an explosion energy of $\approx5\times10^{51}\,\mathrm{erg},$ and a $^{56}$Ni mass of $\approx1.7$ \Msun{}. 

\begin{figure}
    \centering
\includegraphics[width=0.45\textwidth]{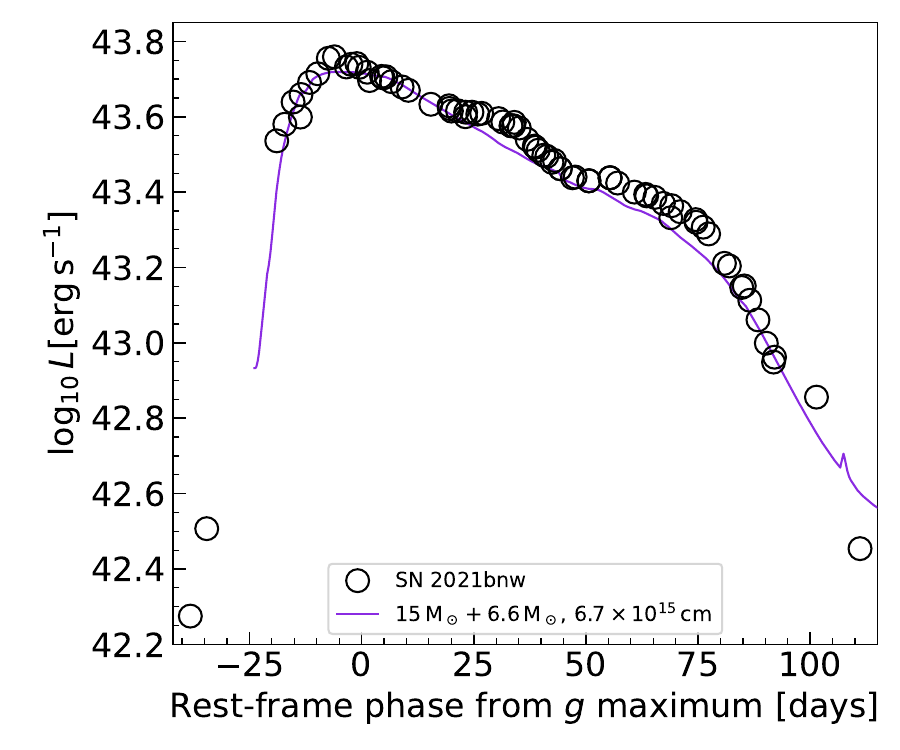}
    \caption{Pseudo-bolometric LC of SN~2021bnw (open circles) compared with the \texttt{STELLA} model \texttt{he90d5he1m6X1W1} (solid purple line).}
    \label{fig:stella1}
\end{figure}

We show in Fig.~\ref{fig:stella1} the \texttt{STELLA} model \texttt{he90d5he1m6X1W1} computed by Kozyreva et al. for a $^{56}$Ni+He-rich CSM scenario. This model was computed assuming $M(^{56}{\rm Ni})=1.7\,{\rm M}_\odot$, $M_{\rm ejecta}=15\,{\rm M}_\odot$, $M_{\rm CSM}=6.5\,{\rm M}_\odot$, and a kinetic energy $E_{\rm kin}=3.8\times10^{51}\,\mathrm{erg}$. 
While the \texttt{STELLA} LC is able to capture the maximum/post-maximum epochs of the pseudo-bolometric LC of SN~2021bnw, it does not predict the two initial detections. Further investigation of these early-time observations, as well as of the progenitor properties of SN~2021bnw within a single-star evolutionary framework, are presented in \citet{kozyrevaetal2026}. We mention that the choice of $^{56}$Ni as the central heating source is not supported by specific observational evidence and might be replaced by an alternative energy-injection mechanism, such as the spin-down radiation lost by a newly born magnetar \citep[see][for the application of this scenario to the case of SN~2021bnw]{poidevinetal2023}. 

We caution that the $^{56}$Ni mass and the kinetic energy required by the \texttt{STELLA} models imply a significant departure from an ordinary core collapse (see Introduction), motivating the consideration of alternative explosion channels, such as the collapsar scenario \citep[e.~g.][]{macfadyenandwoosley1999} or magneto-rotational core collapse \citep[e.~g.][]{winteleretal2012,moestaetal2018,reichertetal2023}. In principle, a PISN is also able to yield much more $^{56}$Ni than an ordinary core collapse, but PISN theory predicts a very slowly evolving LC, which does not match the observed LC of SN~2021bnw.

While we are not able to draw any conclusions in this regard, we note that under the assumption that CSM interaction played a significant role in shaping the LC of SN~2021bnw, the absence of its clear spectroscopic signatures \citep[see e.~g.][]{smithetal2010} in the spectra of SN~2021bnw (and in general, of SLSNe I) might indicate asymmetry in the ejecta–CSM configuration \citep[e.~g.][]{smith2017,andrewsandsmith2018}. Such asymmetries arise naturally in explosion scenarios involving strong rotation and magnetic fields, as expected in the aforementioned alternative core-collapse pathways.
\section{Conclusions}
\label{sec:conclusions}
The supernova SN~2021bnw joins the number of SLSNe I that exhibit helium lines in their spectra, otherwise known as SLSNe Ib. We have presented the outcome of the extensive optical/NIR spectro-photometric follow-up campaign we led for it. Specifically, its $g$-filter LC reaches an absolute peak magnitude of $-20.7$ mag, placing it on the faint wing of the SLSN-I luminosity function. The optical spectra of SN~2021bnw follow the typical spectral evolution of SLSNe I, and the late NIR spectrum, taken 87 days after maximum luminosity, revealed $\lambda\,1.1\,\mu{\rm m}$ and $\lambda\,2\,\mu{\rm m}$ He features (see Sec.~\ref{sec:spectroscopy}). We further investigated and confirmed this identification with the radiative-transfer software \texttt{TARDIS}. Based on the step-like abruptly evolving LC, we ascribed the He spectral features to the interaction of the SN ejecta with H-devoid and He-rich CSM, as was suggested for its sibling SLSN I SN~2017egm. We therefore modeled the LC of SN~2021bnw with the hydrodynamics radiative-transfer code \texttt{STELLA} as a combination of two events: $i)$ the ejection of $6-7$~\Msun{} of a H-free and He-rich shell of the outermost layers of a massive star prior to the terminal explosion, and $ii)$ the actual SN explosion with an energy of $5\times10^{51}\,\mathrm{erg}$, yielding 1.7~\Msun{} of $^{56}$Ni and ejecting $15-20\,\mathrm{M_\odot}$ after the core collapse \citep{kozyrevaetal2026}. However, we were not able to confirm these estimates against a thorough stellar evolution study, and we note that a binary-progenitor scenario might also provide a viable explanation for SN~2021bnw.
\section*{Data availability}
Table~\ref{tab:opt_tables} is only available in electronic form at the CDS via anonymous ftp to \texttt{cdsarc.u-strasbg.fr} (\texttt{130.79.128.5}) or via \texttt{http://cdsweb.u-strasbg.fr/cgi-bin/qcat?J/A+A/}. Spectra are available via Wiserep at \texttt{https://www.wiserep.org/object/17556}.

\begin{acknowledgements}
 We thank Ósmar Rodríguez and the anonymous referee for their valuable comments that improved our work. A. F. acknowledges funding by the European Union – NextGenerationEU RFF M4C2 1.1 PRIN 2022 project ``2022RJLWHN URKA'' and
by INAF 2023 Theory Grant ObFu 1.05.23.06.06 ``Understanding R-process \& Kilonovae Aspects (URKA)''. A. R. acknowledges financial support from the GRAWITA Large Program Grant (PI P. D’Avanzo) and from the PRIN-INAF 2022 "Shedding light on the nature of gap transients: from the observations to the models". T.E.M.B. is funded by Horizon Europe ERC grant no. 101125877. This work makes use of observations from the Las Cumbres Observatory network. MN is supported by the European Research Council (ERC) under the European Union’s Horizon 2020 research and innovation program (grant agreement No.~948381). Y.-Z. Cai is supported by the National Natural Science Foundation of China (No. 12303054), the Yunnan Fundamental Research Projects (Grant Nos. 202401AU070063, 202501AS070078), the National Key Research and Development Program of China (Grant No. 2024YFA1611603), and the International Centre of Supernovae, Yunnan Key Laboratory (No. 202302AN360001). Y.-Z. Cai and A. R. also acknowledge financial support from the SOXS project (PI S. Campana). C. P. G. acknowledges financial support from the Secretary of Universities and Research (Government of Catalonia) and by the Horizon 2020 Research and Innovation Program of the European Union under the Marie Sk\l{}odowska-Curie and the Beatriu de Pin\'os 2021 BP 00168 program, from the Spanish Ministerio de Ciencia e Innovaci\'on (MCIN) and the Agencia Estatal de Investigaci\'on (AEI) 10.13039/501100011033 under the PID2023-151307NB-I00 SNNEXT project, from Centro Superior de
Investigaciones Cient\'ificas (CSIC) under the PIE project 20215AT016 and the program Unidad de Excelencia Mar\'ia de Maeztu CEX2020-001058-M, and from the Departament de Recerca i Universitats de la Generalitat de Catalunya through the 2021-SGR-01270 grant. N. E. R. acknowledges support from the PRIN-INAF 2022, ``Shedding light on the nature of gap transients: from the observations to the models'' and from the Spanish Ministerio de Ciencia e Innovaci\'on (MCIN) and the Agencia Estatal de Investigaci\'on (AEI) 10.13039/501100011033 under the program Unidad de Excelencia Mar\'ia de Maeztu CEX2020-001058-M. T.-W. C. acknowledges the financial support from the Yushan Fellow Program by the Ministry of Education, Taiwan (MOE-111-YSFMS-0008-001-P1) and the National Science and Technology Council, Taiwan (NSTC grant 114-2112-M-008-021-MY3). 
P. B. is supported by the grant RSF 24-12-00141 for modeling supernova light curves with the \texttt{STELLA} code. This work makes use of observations from the Las Cumbres Observatory network (LCO). The LCO team is supported by NSF grants AST-1911225 and AST-1911151. This research made use of \texttt{TARDIS}, a community-developed software package for spectral synthesis in supernovae \citep{kerzendorfandsim2014}. The development of \texttt{TARDIS} received support from GitHub, the Google Summer of Code initiative, and from ESA's Summer of Code in Space program. \texttt{TARDIS} is a fiscally sponsored project of NumFOCUS. \texttt{TARDIS} makes extensive use of Astropy and Pyne. Based on observations collected at the European organization for astronomical research in the Southern Hemisphere, Chile, as part of ePESSTO+ (the advanced Public ESO Spectroscopic Survey for Transient Objects). Based on observations collected at the European Organization for Astronomical Research in the Southern Hemisphere, Chile, as part of ePESSTO+ (the advanced Public ESO Spectroscopic Survey for Transient Objects Survey – PI: Inserra). ePESSTO+ observations were obtained under ESO program IDs 1103.D-0328, 106.216C and 108.220C.. The data presented here were obtained in part with ALFOSC, which is provided by the Instituto de Astrofisica de Andalucia (IAA) under a joint agreement with the University of Copenhagen and NOT.
Based on observations collected at Copernico and Schmidt telescopes (Asiago, Italy) of the INAF-Osservatorio Astronomico di Padova. Based on observations made with the Gran Telescopio Canarias (GTC), installed in the Spanish Observatorio del Roque de los Muchachos of the Instituto de Astrofísica de Canarias, in the island of La Palma. Based on observations obtained with the Samuel Oschin Telescope 48-inch and the 60-inch Telescope at the Palomar Observatory as part of the Zwicky Transient Facility project. Based on observations obtained with the Samuel Oschin Telescope 48-inch and the 60-inch Telescope at the Palomar Observatory as part of the Zwicky Transient Facility project. ZTF is supported by the National Science Foundation under Grant No. AST-2034437 and a collaboration including Caltech, IPAC, the Weizmann Institute of Science, the Oskar Klein Center at Stockholm University, the University of Maryland, Deutsches Elektronen-Synchrotron and Humboldt University, the TANGO Consortium of Taiwan, the University of Wisconsin at Milwaukee, Trinity College Dublin, Lawrence Livermore National Laboratories, IN2P3, University of Warwick, Ruhr University Bochum, Cornell University, and Northwestern University. Operations are conducted by COO, IPAC, and UW. SED Machine is based upon work supported by the National Science Foundation under Grant No. 1106171. The Gordon and Betty Moore Foundation, through both the Data-Driven Investigator Program and a dedicated grant, provided critical funding for SkyPortal \citep{vanderwaltetal2019,coughlinetal2023}.

\end{acknowledgements}
\bibliographystyle{aa}
\bibliography{reference.bib}
\begin{appendix}
\onecolumn
\section{Observations and data reduction}
\subsection{Photometry}
\label{app:photo}
Optical/NIR photometry of SN~2021bnw was obtained exploiting several observational facilities: 
$i)$ New Technology Telescope (NTT) with the EFOSC2 \citep[ESO Faint Object Spectrograph and Camera,][]{buzzonietal1984} and SOFI \citep[Son of ISAAC,][]{moorwoodetal1998} Camera, La Silla Observatory, Chile via ePESSTO+; 
$ii)$ the 2.56m Nordic Optical Telescope (NOT) via the NOT Unbiased Transient Survey \citep[NUTS2, ][]{holmboetal2019} + the Alhambra Faint Object Spectrograph and Camera (ALFOSC), La Palma Observatory, Spain;
$iii)$ with the camera Sinistro built for the 1-m class telescopes of the Las Cumbres Observatory\footnote{\texttt{http://lco.global}} (LCO) telescope network under the Global Supernova Project (GSP) collaboration; 
$iv)$ the 1.82m Copernico+AFOSC and Schmidt telescopes operated by INAF Osservatorio Astronomico di Padova at Asiago-Cima Ekar. In addition, we used the ATLAS \citep[Asteroid Terrestrial-impact Last Alert System, ][]{tonryetal2018a,tonryetal2018} $o$-, $c$-filters photometry and the ZTF $g$- and $r$-filters photometry. ZTF uses the Samuel Oschin 48-inch (1.22 \,m) Schmidt telescope at Palomar Observatory on Mount Palomar (USA), utilizing the Fritz Marshal System \citep{vanderwaltetal2019,coughlinetal2023}. The Samuel Oschin Schmidt telescope is equipped with a 47-square-degree camera \citep{dekanyetal2020} and monitors the entire northern hemisphere every 2--3 days in the $g$ and $r$ bands to a depth of $\sim20.7$~mag ($5\sigma$; \citealt{bellmetal2019a,grahametal2019}) as part of the public ZTF Northern Sky Survey \citep{bellmetal2019b}. We retrieved the host-subtracted photometry via the Infrared Processing and Analysis Center ZTF forced-photometry service \citep{mascietal2023a}. This service uses the data-reduction techniques outlined in \citet{mascietal2019a}. We cleaned and calibrated the data following \citet{mascietal2023a}.

We performed magnitude measurements with the \texttt{ecsnoopy} package\footnote{\texttt{ecsnoopy} is a python package for SN photometry using PSF fitting and template subtraction developed by E. Cappellaro. A package description can
be found at \texttt{http://sngroup.oapd.inaf.it/}.} \citep{cappellaro2014}. Background contamination was removed from the SN light by performing template subtraction with the \texttt{hotpants} tool \citep{becker2015}. $u$-, $g$-, $r$-, $i$-, $z$-filters templates were obtained via the Sloan Digital Sky Survey (SDSS) and $B$-, $V$-filter templates were observed at the NOT telescope on 2023 May 5, that is, 833 days after maximum. We assume that at this epoch SN~2021bnw faded well below the detection limit. $J$-, $H$-, $K$-filter templates were obtained via the Two Micron All Sky Survey \citep[2MASS, ][]{skrutskieetal2006}. Then, we calibrated the instrumental $u$-, $g$-, $r$-, $i$-, $z$- and $J$-, $H$-, $K$-filters magnitudes against a series of non-saturated field stars present in the Pan-STARRS \citep[Panoramic Survey Telescope and Rapid Response System][]{chambersetal2016} and the 2MASS catalogs, respectively, while the $B$-, $V$-filters magnitudes were calibrated following the color transformations of \citet{chonisandgaskell2008}. The reduced optical and NIR photometry is listed in Tables~\ref{tab:opt_tables}, \ref{tab:nir_tables}.
\subsection{Spectroscopy}
\label{app:spectra}
We obtained optical spectroscopy of SN~2021bnw at NTT+EFOSC2 via ePESSTO+, at NOT+ALFOSC via NUTS2, at 1.82m Copernico+AFOSC at INAF, Osservatorio Astronomico di Padova at Asiago-Cima Ekar and LCO 1.0m telescopes+Sinistro and at the Palomar 60-inch telescope+SEDM \citep[Spectral Energy Distribution Machine, ][]{blagorodnovaetal2018}. We also obtained a NIR spectrum at 87 days after maximum luminosity at Keck+NIRES \citep[Near-Infrared Echellette Spectrometer,][]{wilsonetal2004}. The log of the spectroscopic observations is shown in Table~\ref{tab:speclog}.

Raw optical NTT+EFOSC2 spectroscopic frames were initially corrected for overscan, bias and flat field. Then the extracted trace was calibrated in wavelength with HeAr (for the NTT+EFOSC2), HeNe (for NOT+ALFOSC spectra), NeHgCd (for the 1.82m- Copernico+AFOSC spectra) lamps and in flux using the spectrum of a spectrophotometric standard star observed on the same night and with the same instrumental set-up of each scientific observation. Moreover, we account for the cosmic-ray contamination with the \texttt{python} package \texttt{lacosmic}\footnote{\texttt{https://lacosmic.readthedocs.io/} .}, based on the Laplacian edge detection (\texttt{L. A. Cosmic}) algorithm by \citet{vandokkum2001}. NTT+EFOSC2 spectra were processed with \texttt{IRAF} \citep[Image Reduction and Analysis Facility,][]{tody1986,tody1993}. For Asiago and NOT spectra, the same procedures were called via the graphical user interface \texttt{foscgui}\footnote{\texttt{foscgui} is a \texttt{python}/\texttt{pyraf} based graphic user interface aimed at extracting
SN spectroscopy and photometry obtained with FOSC-like instruments. It was developed by E. Cappellaro. A package description can be found at \texttt{http://sngroup.oapd.inaf.it/foscgui.html} .}. SEDM spectra are reduced with the tool \texttt{pysedm} \citep{rigaultetal2019}.
\section{Tables}
\onecolumn
\begin{longtable}{llllllll|l}
\caption{Optical magnitudes measurements for SN~2021bnw.}\\
\hline\hline
\label{tab:opt_tables}

MJD&$u$&$B$&$g$&$V$&$r$&$i$&instrument&$\log_{10}L_{\rm bol}$\\
\hline
\endfirsthead
\caption{continued.}\\
\hline\hline
MJD&$u$&$B$&$g$&$V$&$r$&$i$&instrument&$\log_{10}L_{\rm bol}$\\
\hline
\endhead
\hline
\endfoot
59219.44&-&-&$19.69^*$&-&-&-&ZTF&-\\
59219.48&-&-&-&-&$19.82^*$&-&ZTF&-\\
\ldots&\ldots&\ldots&\ldots&\ldots&\ldots&\ldots&\ldots&\ldots\\
\hline
\end{longtable}
\tablefoot{
Magnitudes are AB ($B$, $V$ in Vega), uncorrected for $K$ correction or Galactic extinction. Errors are in parentheses; asterisks denote detection limits. Full table at CDS.
}
\begin{table*}[h!]
\caption{$z$, $J$, $H$, $K$ magnitudes measurements.}
\label{tab:nir_tables}
\centering
\begin{tabular}{llllll}
\hline
\hline
MJD&$z$&$J$&$H$&$K$&instrument\\
\hline
59264.08&-&17.27(0.25)&-&-&NOTCAM\\
59264.09&-&-&-&16.69(0.29)&NOTCAM\\
59278.23&-&17.42(0.43)&-&-&SOFI\\
59278.24&-&-&-&$17.81^*$&SOFI\\
59278.26&-&-&15.92(0.46)&-&SOFI\\
59291.90&$17.01^*$&-&-&-&ALFOSC\\
59308.07&-&17.26(0.03)&-&-&SOFI\\
59308.09&-&-&-&$16.74^*$&SOFI\\
59308.11&-&-&16.0(0.33)&-&SOFI\\
59315.00&-&17.18(0.18)&-&-&SOFI\\
59315.02&-&-&-&$17.0^*$&SOFI\\
59315.03&-&-&16.23(0.06)&-&SOFI\\
\hline
    \end{tabular}
\tablefoot{
$z$ is AB , $J$, $H$, $K$ in Vega, uncorrected for $K$ correction or Galactic extinction. Errors are in parentheses; asterisks denote detection limits.
}
\end{table*}
\begin{table*}[h!]
\caption{Log of the spectroscopic observations and $K$ corrections in $B$, $g$, $V$, $r$, $i$ filters computed on each spectrum.}
\label{tab:speclog}
\centering
    \begin{tabular}{llllllll}
\hline
\hline
MJD&Instrumental set up&Resolution[\AA{}]&$K_B$&$K_g$&$K_V$&$K_r$&$K_i$\\
\hline
59249.29&NTT+EFOSC2&18.3&-0.03&-0.04&0.00&0.00&-0.00\\
59251.57&1.0m+Sinistro&14.7&-0.03&-0.04&0.03&0.05&0.04\\
59255.24&NTT+EFOSC2&-&-0.02&-0.04&-0.00&0.01&0.00\\
59260.13&NOT+ALFOSC&18.8&-0.02&-0.05&-0.01&0.08&0.06\\
59263.03&1.82m+AFOSC&-&-0.05&-0.06&-0.03&-0.03&-0.01\\
59276.21&NTT+EFOSC2&-&-0.04&-0.06&-0.06&-0.05&-0.01\\
59286.17&NTT+EFOSC2&-&-0.14&-0.16&-0.15&-0.16&-0.02\\
59288.87&1.82m+AFOSC&-&-0.12&-0.13&-0.08&-0.10&-0.07\\
59292.40&1.0m+Sinistro&-&-0.13&-0.14&-0.09&-0.07&-0.07\\
59296.13&NTT+EFOSC2&-&-0.05&-0.06&-0.07&-0.05&-0.04\\
59303.44&1.0m+Sinistro&-&-0.06&-0.02&-0.05&-0.11&-0.08\\
59312.90&1.82m+AFOSC&16.7&-0.13&-0.12&-0.07&-0.12&0.52\\
59322.86&NOT+ALFOSC&14.4&-0.19&-0.17&-0.07&0.07&-0.03\\
59333.45&1.0m+Sinistro&-&-0.20&-0.16&-0.16&-0.13&-0.11\\
59337.37&1.0m+Sinistro&-&-0.14&-0.10&-0.10&-0.11&-0.09\\
59338.92&NOT+ALFOSC&14.6&-0.21&-0.19&-0.12&-0.15&-0.09\\
59347.00&SEDM&-&-0.31&-0.23&-0.18&-0.18&-0.11\\
59348.90&NOT+ALFOSC&18.1&-0.24&-0.21&-0.15&-0.18&-0.15\\
59359.37&1.0m+Sinistro&21.9&-0.19&-0.05&-0.04&-0.12&-0.13\\
59362.88&NOT+ALFOSC&14.9&-0.27&-0.23&-0.15&-0.13&-0.08\\
59364.00&Keck+NIRES&-&-&-&-&-\\
59370.34&1.0m+Sinistro&15.0&-0.37&-0.23&-0.09&-0.15&-0.25\\
59375.93&NOT+ALFOSC&16.7&-0.33&-0.28&-0.18&-0.20&-0.17\\
\hline
    \end{tabular}
\end{table*}

\pagebreak

\end{appendix}

\end{document}